\documentclass[aps,amsmath,pre,floatfix,superscriptaddress]{revtex4-2}
\usepackage{graphicx}
\usepackage{geometry}
\usepackage{wrapfig}
\usepackage{comment}
\usepackage{amsmath}
\usepackage[export]{adjustbox}
\usepackage{float}
\usepackage{amssymb}
\usepackage[utf8]{inputenc}
\usepackage{afterpage}
\usepackage{lipsum}  
\usepackage{appendix}
\usepackage{natbib}

\usepackage{hyperref}% add hypertext capabilities
\hypersetup{colorlinks=true, linktoc=all, linkcolor=blue, linktocpage, citecolor=blue}

\usepackage{geometry}\usepackage{geometry}

\usepackage[normalem]{ulem}

\begin{document}
\title{Evolution of commitment in the spatial Public Goods Game \\ through institutional incentives}
\author{Lucas S. Flores }
\address{Instituto de Física, Universidade Federal do Rio Grande do Sul, CP 15051, CEP 91501-970 Porto Alegre - RS, Brazil}
\author{The Anh Han }
\address{School of Computing, Engineering and Digital Technologies, United Kingdom}

\begin{abstract}
    Studying social dilemmas prompts the question of how cooperation can emerge in situations where individuals are expected to act selfishly.
    Here, in the framework of the one-shot Public Goods Game (PGG), we introduce the concept that  individuals can potentially adjust  their behaviour based on the cooperative commitments made by other players in the group prior to the actual PGG interaction.
    To this end, we establish a commitment threshold that  group members must meet for a commitment to be formed. We explore the effects of punishing commitment non-compliant players (those who commit and defect if the commitment is formed)  and rewarding commitment-compliant players (those who commit and cooperate if the commitment is formed). 
    In the presence of commitment and absence of an incentive mechanism, we observe that conditional behaviour based on commitment alone can enhance cooperation, especially when considering a specific commitment threshold value.
   In the presence of punishment, our findings suggest that the survival of cooperation  most likely happen at intermediate  commitment thresholds.  Notably,  cooperation is maximised at high thresholds,  when punishment occurs more frequently.
    We also see that when cooperation rarely survives, a cyclic behaviour emerges, facilitating the persistence of cooperation.
    For the reward case, we found that cooperation is highly frequent regardless of the  commitment threshold adopted.
\end{abstract}

\maketitle

\section{Introduction}

Prior to embarking on a collective project, individuals involved may solicit commitment from group members and estimate how interested they are in contributing to the group's efforts. This assessment helps them determine whether it is worthwhile to  initiate the endeavour  and/or if it would be beneficial to join.
Commitment mechanisms for enhancing cooperation are widespread in nature, which exist  in various  forms and contexts,  including legal contracts and pledges  \cite{nesse2001evolution}, marriage \cite{swensen1985commitment},  deposit-refund schemes \cite{cherry2013enforcing}, emotion-based  \cite{FrankChapter2001} or reputation-based commitment  \cite{frank88,krellner2023importance}. 
Both empirical and theoretical studies demonstrated that high levels of cooperation can be achieved through reliable commitments \cite{ostrom2009understanding,dannenberg2016non,zumbansen2007law,han2022voluntary,chen1994effects}. 
They enable individuals to reach mutual cooperation even when there is little knowledge about others' past behaviours \cite{han2022Interface,han2017evolution,sasaki2015,NBOgbo}, as it requires them to reveal their preferences or intentions \cite{han2015synergy,tomasello2005understanding,Silk2001}.  

However,  prior models of commitment have mainly focused on well-mixed population settings, potentially overlooking the significant influence of commitment dynamics within a population's actual network structure. This structure dictates who may form commitments with whom \cite{Szabo2007,barabasi2014linked}, ultimately determining the global worthiness of arranging conditional commitments.  
Network reciprocity plays a crucial role in shaping human social interactions, fostering the creation of cooperative clusters and thereby influencing cooperation dynamics \cite{Szabo2007,perc2013evolutionary,szolnoki2011phase,barabasi2014linked}. Here, our study reveals that this critical aspect also significantly influences how commitments contribute to the emergence of cooperation.

Moreover, the initiation of many collective projects hinges on the majority of participants making a commitment to contribute towards a  common good. 
For instance, for a cooperative hunting endeavor to take place, it typically requires a sufficient number of participants ready and willing to participate \cite{alvard2002rousseau,stander1992cooperative}.
While some international agreements require ratification by all parties before entering into force, most (especially global treaties) require a minimum of less than the total number of negotiating countries \cite{barrett2003environment,cherry2013enforcing}. In general, it appears that the necessary level of commitment is contingent on the specific nature of the problem at hand.
However, this issue has been under-explored in theoretical modeling, especially in the context of spatial group-interaction settings.

As motivated, herein we investigate
 the potential of a conditionally applied commitment strategy, contingent on the required commitment level from the group, to promote the evolution of cooperation in a structured population. Our analysis is carried out in the context of the one-shot  Public Goods Game (PGG) \cite{hofbauer1998evolutionary,key:Hauert2007,ostrom2019public, sigmund2010social,dreber2008winners}.  
 In this game,  a group of $G$ players have the choice to invest (cooperate, paying a cost $c$) or not (defect, paying nothing) in a common pool of the group. All their contributions are then multiplied by a factor $r \ (1 < r < G$) and in the end the result is divided equally among all players, regardless of their initial choice.
Before engaging in a PGG game, players can choose whether or not to join a commitment and cooperate in the game. The commitment is formed if a threshold $\tau$ ($0 \leq \tau \leq G$) regarding the number of committed players is met. Subsequently, players make decisions in the game based on whether the commitment is formed or not.

In addition, for understanding optimal  incentive mechanisms that  enable  commitment compliance \cite{han2022Interface}, we assess the comparative effectiveness of institutional reward and punishment in promoting cooperation, given a commitment formation threshold. 
Our findings indicate that both punishment and reward mechanisms can positively impact cooperation. Furthermore, even in the absence of these mechanisms, the potential for a cooperator to switch to defection can still contribute to cooperation, especially with a specific commitment threshold.
Notably, in the presence of punishment, we observe that high punishment and high commitment thresholds are  most effective. However, intermediate thresholds can sustain cooperation for difficult PGGs (i.e. those with small values of $r$).
In the case of reward, high levels of cooperation can be achieved regardless of the commitment threshold.

\section{Model}
Players interact following the one-shot PGG of size $G$. First, each player decides to commit or not to cooperate. If the total number of committed players exceeds a given threshold, denoted by $\tau$, the commitment is formed. 
In addition, players can choose to cooperate or not if the commitment is formed, contributing to the common pool of the group.
We establish the possibility that players who committed but then defect may have to pay a fine and those who committed and cooperated may receive a reward. 
If the commitment is not formed (when the commitment threshold is not reached), players can decide to change their choice in the PGG (compared to what they would have played if the commitment was formed).

We denote the steps above as follows: $i$ the decision to accept ($i=A$) or not ($i=N$) to commit; $j$ the decision to cooperate ($j=C$) or not ($j=D$) if the commitment is formed; and $k$ the decision to cooperate ($k=C$) or not ($k=D$) if the commitment is not formed.
For example, the $NCD$ strategy does not commit ($i=N$), cooperates if the commitment is formed ($j=C$), and defects if it is not formed ($k=D$). In total, there are eight possible strategies, see Table \ref{table:eight-strategies}.

\begin{table}
\begin{tabular}{ p{2cm}|p{2.3cm}p{3.3cm}p{3.cm}  }
 \hline
Strategies &  Accept commitment?  & Cooperate in presence of commitment? & Cooperate in absence of commitment? \\
 \hline
 ACC   & Yes    &  Yes  &  Yes \\
 ACD&   Yes  & Yes   &No\\
 ADC &Yes & No&  Yes\\
 ADD    &Yes & No&  No\\
 NCC&   No  & Yes&Yes\\
 NCD& No  & Yes   &No\\
 NDC& No  & No&Yes\\
  NDD& No  & No&No\\
 \hline
\end{tabular}
\caption{The eight strategies with commitment formation.}
\label{table:eight-strategies}
\end{table}

For one group $X$, an individual with strategy $ijk$ has the following payoff
  \begin{equation}
  \label{eq:general_payoff}
      \Pi_{ijk} = \frac{r}{G} \, \sum_{x \in X} c_x  - c_{ijk} + incentive,
      \end{equation}
where  $c_x$ is the contribution from group member $x$ (including the focal player $ijk$),  $c_{ijk}$ is the contribution from the focal player, and the incentive is a reward gained or a punishment incurred by the focal player (explained below, see also Table \ref{table:incentives}).
A player  contributes to the common pool, i.e. $c_{ijk} = c$, if they cooperate when the commitment is formed ($j = C$) and the commitment was actually formed, or if they cooperate when the commitment is not formed ($k = C$)  and the commitment was actually not formed.  Otherwise, the player does not contribute. In that case,  $c_{ijk} = 0$.
Without loss of generality, we set  $c = 1$ in our analysis.

The total budget for providing incentives, i.e., punishment against commitment non-compliant players (i.e., $ADC$ and $ADD$) or reward for commitment-compliant players (i.e., $ACC$ and $ACD$)  is given by $G \delta$ per group, where $\delta$ is the per capita incentive.
This budget $G \delta$  is then divided into two parts based on a relative weight $\omega$ with $0 \leq \omega \leq 1$. 
The reward part $\omega G \delta$ is equally shared among $n_{com}$ committed  compliant players, who received a reward $ \omega G \delta / n_{com}$. 
The punishment part  $(1-\omega) G \delta$ is equally impacted by  $n_{non-com}$ committed non-compliant players, who incur a punishment $ (1-\omega) G \delta / n_{non-com}$. We summarise all incentives in Table \ref{table:incentives}
\begin{table}
\begin{tabular}{ p{2cm}|p{3.3cm}  }
 \hline
Strategies &  incentives \\
 \hline
 ACC   &  $\omega G \delta / n_{com} $   \\
 ACD&   $ \omega G \delta / n_{com} $ \\
 ADC &  $- (1-\omega) G \delta / n_{non-com}$\\
 ADD    & $ - (1-\omega) G \delta / n_{non-com}$ \\
 NCC&   0  \\
 NCD& 0 \\
 NDC& 0  \\
  NDD& 0 \\
 \hline
\end{tabular}
\caption{The eight strategies with commitment formation and their incentives if the commitment is formed. If the commitment is not formed, no incentives is provided to any strategy.}
\label{table:incentives}
\end{table}

A player's accumulated payoff is the sum of payoffs it obtained from the interactions in all groups it participates in (in  case of a square lattice, it receives payoffs from five groups). 
Note that for $\tau=\delta=0$ we have the classical PGG.

In an evolutionary step, first a random player ($ijk$) is selected from the population. Its payoff is  calculated  according to  Equation \ref{eq:general_payoff}. Then, a random neighbor ($i'j'k'$) of ($ijk$) is selected, and we repeat the same calculation for its payoff.
Player ($ijk$) will adopt ($i'j'k'$) strategy according to a probability given by the Fermi update rule,  
\begin{equation} \label{eq.transition}
     W_{ijk \rightarrow i'j'k'}= \frac{1}{1+e^{-(\Pi_{i'j'k'}-\Pi_{ijk})/K}} \,\,,   
\end{equation}
where $K$ is a noise related to irrationality.
It is important to notice that commitment coevolves with cooperation and defection.

This evolutionary step is repeated $N$ times where $N$ is the population size, characterising one Monte Carlo step (MCS). We set the total simulation time as $t = 10^5$ MCS which is enough for the equilibrium to be reached.
In line with previous works \cite{Szabo2007,Rand2011,perc2013evolutionary,perc2017statistical}, we perform our simulations on a square lattice with von Neumann neighborhood, with size $N = 100^2$ and $K=0.1$. We initialise the population with an equal number of each strategy.
Since each player can cooperate or not according to their neighborhood, we define the cooperation frequency as the number of  times that an interacting player cooperates per group.

\section{Results}
As the baseline, we begin by exploring the effect of commitment on the evolution of cooperation in the spatial PGG in the  absence of any incentive mechanism (Section A).
We then study the impact of  punishment of commitment non-compliant behaviour (Section B) and  reward of commitment-compliant behaviour (Section C). 

\subsection{Commitment without incentives ($\delta=0$)}

\begin{figure}
    \centering
    \includegraphics[width=0.5\columnwidth]{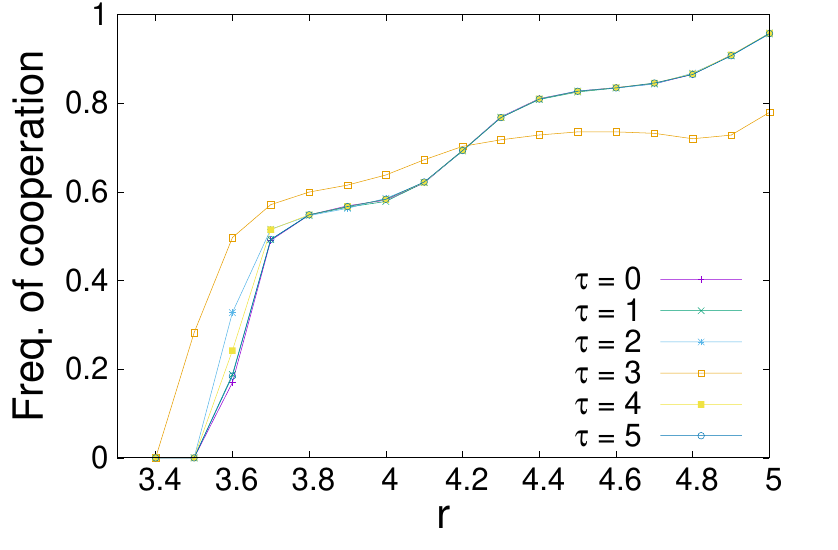}     
    \caption{Frequency of cooperation as a function of the public goods multiplication factor, $r$. For all $\tau \neq 3$, a similar outcome is observed to that of the classical PGG game ($\tau = 0$).    
    For $\tau < 3$, only non-committers survive whereas for $\tau > 3$,  only committers do. For $\tau = 3$, a coexistence of committers and non committers,  allowing conditional strategies ($ACD$ and $NDC$) to gain an advantage over unconditional defectors ($NDD$ and $ADD$).}
    \label{classical}
\end{figure}

We show in Fig.~\ref{classical} the cooperative density for varying $r$ for all threshold values.
For $\tau = 0$, we reproduce the   the classical PGG (i.e., without commitment), where every player has a fixed choice in the PGG  since the commitment is always formed.
For all other thresholds, except $\tau = 3$, we observe a similar cooperation outcome to the classical case.   
An important remark is that spatial reciprocity plays a crucial role in games in structured populations. In such games,  cooperators can form clusters to survive, by avoiding defecting neighbours. We notice that only some commitment-based strategies (Table \ref{table:eight-strategies}) have spatial reciprocity, meaning that when they cluster together they will cooperate with each other. They are $NDC$, $NCC$, $ACC$, and $ACD$. 
Nevertheless, if the commitment threshold is low and thus easily formed, $NDC$ players would have an advantage over the committing strategies above, since they can cluster and cooperate with each other while at the same time defecting against them. 
Moreover, they are not exploitable by defectors that commit, because they would defect in such players' presence. Note that, unlike $NDC$,  $NCC$ possesses neither of those two advantages.
Thus, defectors that do not commit have a benefit and cooperators that commit or have unconditional strategies are exploitable. 
On the other hand, if the commitment threshold is high, it would be harder to be formed, and therefore $ACD$ would have spatial reciprocity at the same time exploiting the other strategies that also have it. They are not exploited by defectors who do not commit since they would defect in their presence.
Thus, defectors that commit have a benefit and cooperators that do not commit or have unconditional strategies are exploitable. 

In summary, low thresholds select for the non-committing strategies ($i = N$), while high thresholds for the committing ones ($i = A$). 
Thus, strategies will always cooperate or defect unconditionally, resulting in the classical PGG.
%
%That is, in this case we have the same classical scenario but with cooperators and defectors having different labels.

Interestingly, for $\tau = 3$, we have a different scenario from the classical game, even without punishment or reward. We observe  that non-committing strategies ($NDD$ and $NDC$) coexist with committing ones ($ADD$ and $ACD$).
The intermediate threshold $\tau=3$ is high enough for $ACD$ to exploit $NDC$ (capable of invading a cluster without reaching the commitment thresholds) and low enough for the $NDC$ to exploit the $ACD$ (able to invade a cluster and still meet the commitment). In the end, we see that their densities always converge to the same value when interacting alone, independently of $r$.
Therefore, we end up with committing and non-committing strategies being able to coexist. 
If we set only two strategies at a time in the population, we observe that $ACD$ performs better against $NDD$ than $NDC$.
The same is true for $ADD$, where $NDC$ performs better than $ACD$. 
This can be understood by the fact that, when we have an interaction between $NDD$ and $NDC$, we have the classical scenario, where they never change strategy. When $NDD$ interacts with $ACD$, a cooperator can sometimes defect and avoid exploitation.
Therefore, acting conditionally to avoid exploitation is better than unconditionally cooperating, for low $r$ values. For high $r$ values, unconditional cooperation becomes viable and therefore can outperform defective strategies.

\begin{figure}[H]
    \centering
    \includegraphics[width=0.3\columnwidth]{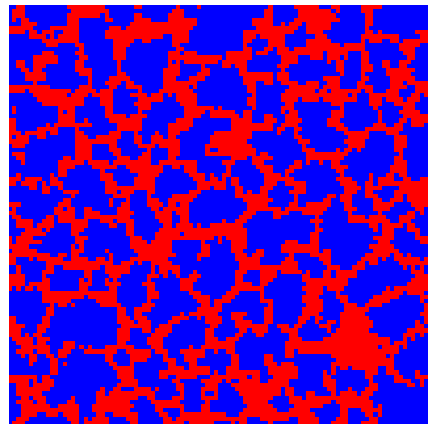}
    \includegraphics[width=0.3\columnwidth]{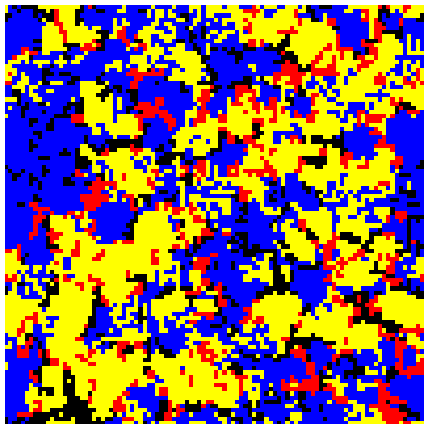}
    \includegraphics[width=0.3\columnwidth]{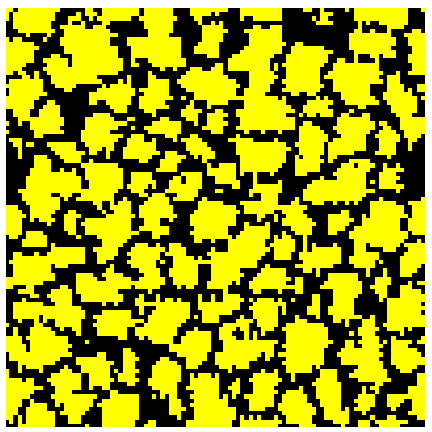}
    \caption{ Snapshots of the population at  equilibrium for $r=4$ and different threshold values: $\tau = 2$ (left), $3$ (middle) and $4$ (right). For $\tau = 2$, only non-committers are present, where $NDC$ players (blue) cluster around $NDD$ (red). For $\tau = 4$ we see that only committers survive, where $ACD$ players (yellow) cluster in a sea of $ADD$ (black). For $\tau=3$ there are not anymore clear clusters and we see that $ACD$ and $NDC$ keep invading one another in the presence of defectors ($NDD$ and $ADD)$.}
    \label{snap_classic}
\end{figure}

We illustrate the observations above with snapshots of the population when the equilibrium is reached, see  Fig.~\ref{snap_classic}. We observe that for low threshold values ($\tau<3$), only non-committers survive by clustering ($NDC$ and $NCC$) in a sea of $NDD$ and $NCD$ players, whereas for high threshold values ($\tau>3$) only committers survive by clustering ($ACD$ and $ACC$) in a sea of $ADC$ and $ADD$ players. 
For $\tau = 3$ there are no longer clearly formed clusters as in the previous cases. This is because $ACD$ and $NDC$ can invade one another.

%%%%%%%%%%%%%%%%%%%%%%%%%%%%%%%%%%%%%%%%%%%%%%%%%%%%%%%%%%%%%%
\subsection{Punishment ($\omega=0$, for varying $\delta$) }

\begin{figure}%[H]
    \centering

    \includegraphics[width=0.3\columnwidth]{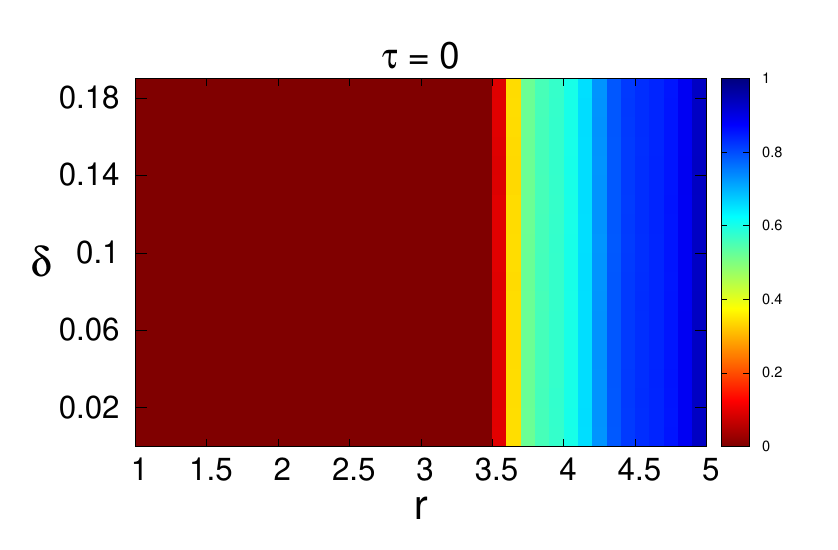}
    \includegraphics[width=0.3\columnwidth]{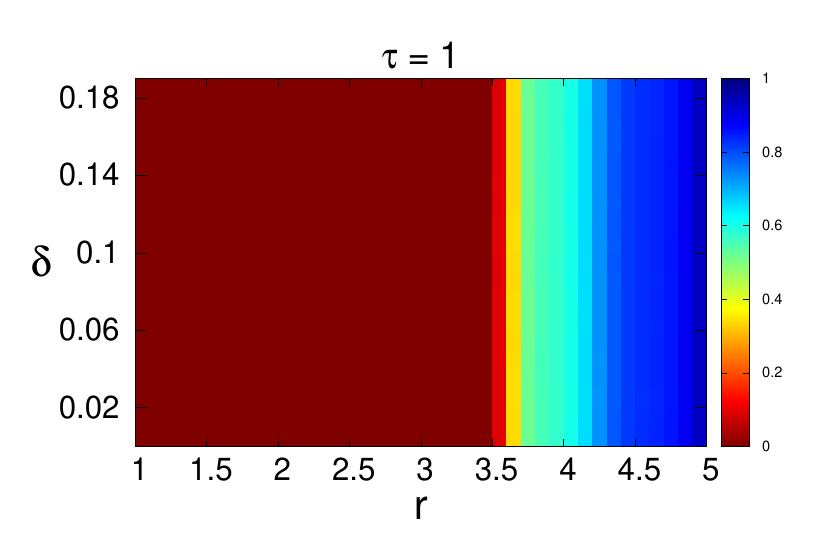}
    \includegraphics[width=0.3\columnwidth]{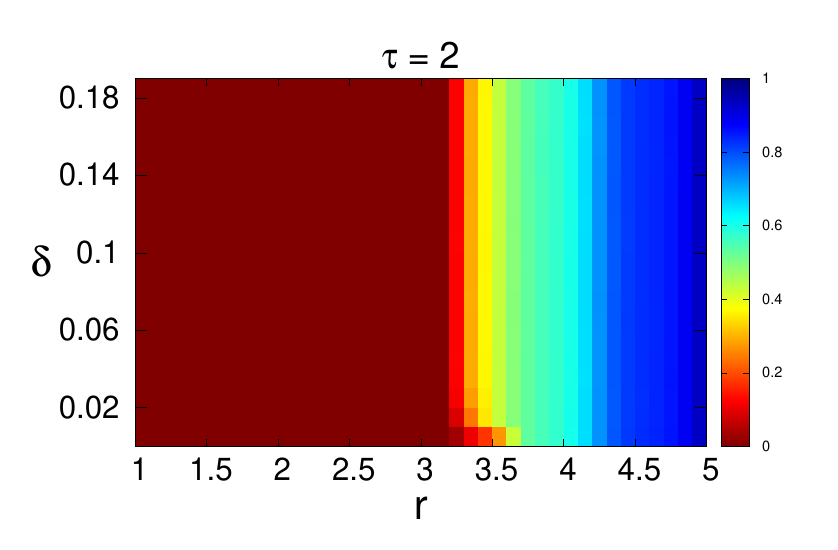}
    \includegraphics[width=0.3\columnwidth]{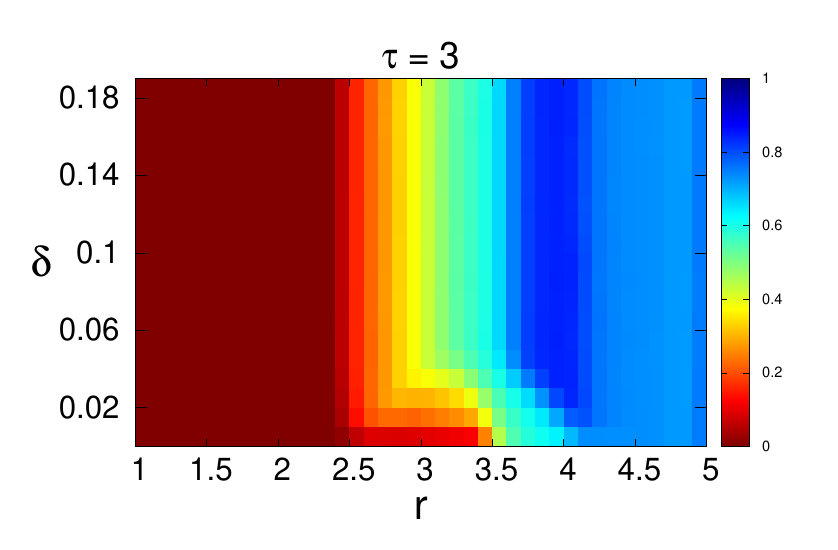}
    \includegraphics[width=0.3\columnwidth]{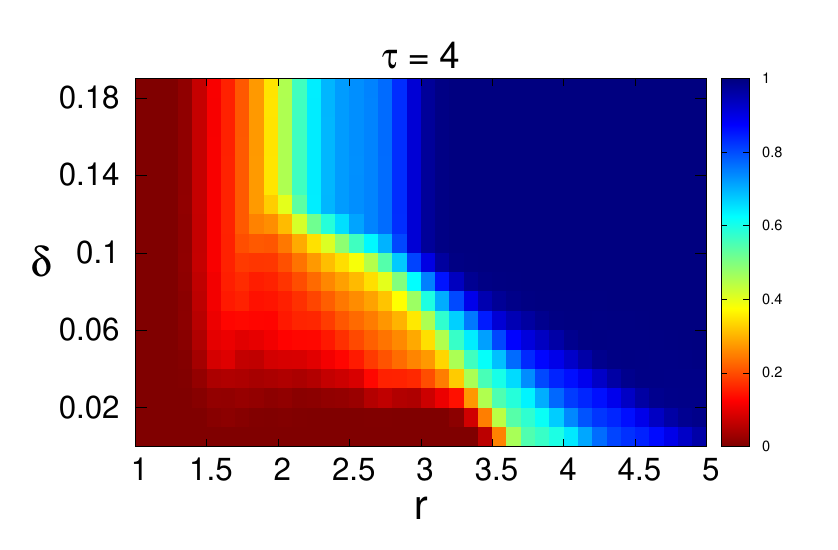}
    \includegraphics[width=0.3\columnwidth]{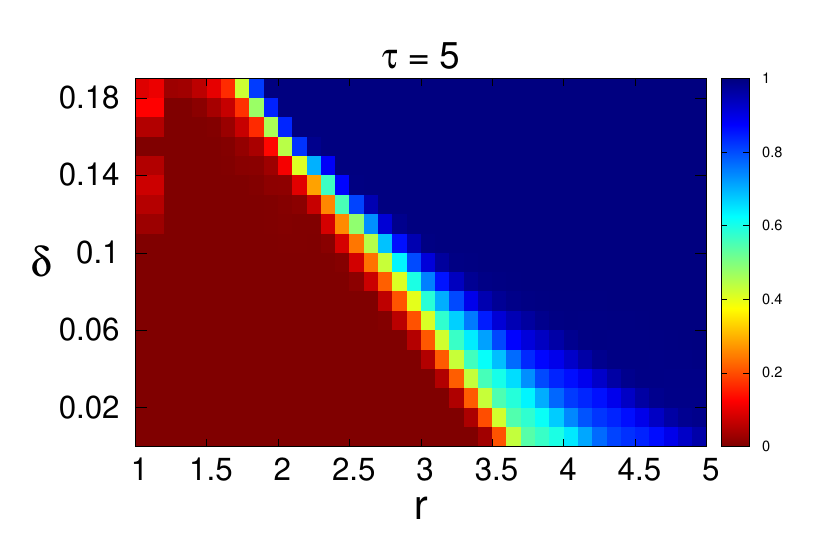}
    \caption{ Phase diagram $r \times \delta$ for the density of cooperative behaviour for all threshold values, in the presence of  punishment. We observe that for $\tau \geq 2$, cooperation  benefits from the presence of punishment, in the sense of a reduced critical $r$ for cooperation to prevail. The effect of stronger punishments stagnate slower for high thresholds, allowing cooperation to survive for lower $r$ values. Despite that, intermediate thresholds (i.e. $\tau = 3, \ 4$) can sustain cooperation for the same $r$ value with a smaller punishment value. Another interesting observation is that for $\tau=5$ cooperation can even survive for very hard PGG (i.e. small $r$,  $1 < r \leq 1.1$), due to a cyclic behaviour.}
     \label{freq_w0}
\end{figure}

Here we explore the effect of punishment, where players who committed to cooperate yet defect after the formation of a  commitment are punished and thus have a decrease in their payoff. 
Indeed, Fig. \ref{freq_w0} shows the frequency of cooperation for different thresholds, in the presence of punishment.
We observe  that, for a sufficiently high threshold, namely, $\tau \geq 2$, punishment leads to improved cooperation compared to the classical scenario (note that for $\tau =0$ the model is equivalent to the classical PGG). For $\tau=1$,  there is no improvement compared to the classical scenario.
This is because in this case a commitment is never formed, therefore no punishment is applied. This results in the same classical scenario since cooperators and defectors who do not commit are possible strategies. 
We observe significant improvements only for high thresholds ($\tau=4$ and $5$),  commitments being  formed and therefore the punishment being applied.

\begin{figure}[H]
    \centering
    \includegraphics[width=0.6\columnwidth]{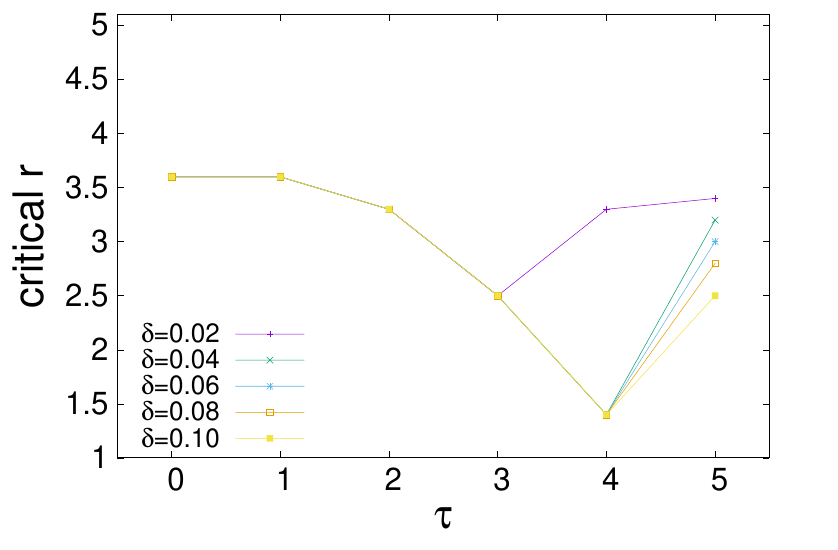}
    \caption{We show the critical value of $r$ that sustains cooperation, as a function of the threshold $\tau$, for different  costs of punishment, $\delta$. 
    We observe that weaker punishments are more effective for intermediate thresholds to sustain cooperation.
    }
    \label{commit}
\end{figure}

We also observe that for all threshold values eventually increasing punishment strength stops to affect cooperation. The lower the threshold, the sooner this behaviour occurs, resulting in  cooperation enhancement in high thresholds and low $r$ values for stronger punishments. Therefore, high threshold values are most conducive to the emergence and dominance of cooperation.
Interestingly, cooperation is not dominant in the population for $\tau = 3$ in our parameters range. This is due to the coexistence between $ACD$ and $NDC$ players. For the lowest $r$ values where cooperation survives, we observe that only $ACD$ players coexist with $NDD$ ones. Increasing $r$  benefits cooperative strategies and therefore, increases cooperative density. But at the same time, the increase of $r$ allows $NDC$ players to survive in the population. Their presence  is detrimental for the overall cooperative density since they are less effective against non-committing defectors, as discussed in the previous section.
Despite that, intermediate thresholds can sustain cooperation for smaller $r$ values if the punishment is low enough. We illustrate this in Fig.~\ref{commit}, where we plot the critical values of $r$ for the survival (for low punishment) of cooperation, as a function of the threshold $\tau$.
In general, cooperation is enhanced in the sense of survival (minimal $r$ that cooperation survives) for intermediate thresholds. Moreover, cooperation is dominant (minimal $r$ that cooperation dominates) when the threshold is high and punishment is strong.

\begin{figure}[H]
    \centering
    \includegraphics[width=0.8\columnwidth]{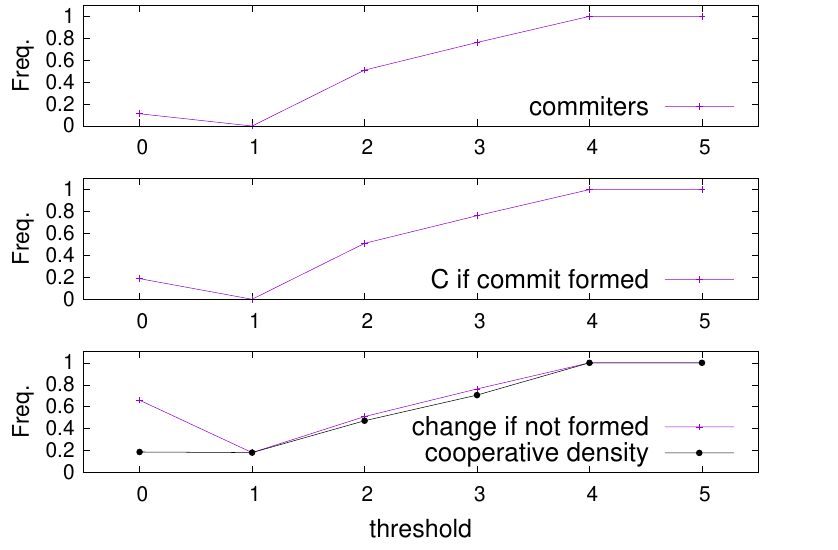}    
    \caption{ Density of strategies for varying the threshold, for $r=3.6$ and $\delta=0.19$. Increasing threshold selects for $ACD$ strategy. We observe that for low thresholds cooperation evolves due to the conditional non-committers, while for high thresholds, it evolves due to the conditional committers. Note that the density of cooperation is slightly smaller than the density of committers because if committers and non-committers coexist the former can sometimes defect.}
    \label{commit2}
\end{figure}

Now, we examine which strategies contribute to the presence of cooperation in Fig.~\ref{freq_w0}.
Fig.~\ref{commit2} shows the frequency of each behaviour as a  function of the threshold for $r=3.6$ and $\delta=0.19$.
The same trend is observed from the one-shot PGG when changing thresholds. For low thresholds, only non-committers survive, while for high thresholds, only committers do. But now, this is a more continuous transition since we observe the coexistence of committers and non-committers for more thresholds, such as 2, 3, and 4 (depending on the $r$ value). 
It can also be seen that for all thresholds, the cooperative density matches the density of those strategies that would change their choice in the PGG if the commitment is not formed. The case $\tau=0$ is an exception because the commitment is always  formed.

Recalling that for the case of commitment without incentives (Section A), we observed that high thresholds selected for defectors that commit and low thresholds for defectors that do not commit. As such, a commitment is likely being formed for high thresholds, thus, defectors are always be punished. This is why high punishments under high thresholds are highly effective (for promoting cooperation). For lower thresholds, a commitment is less likely to be formed, and thus defectors are punished less often. Therefore even high punishments are not highly effective.
Another interesting observation is that for $\tau = 5$, cooperation can even survive for very low values of $r$ (corresponding to  PGG games where it is very hard for cooperation to survive, e.g. $r=1.1$), due to a cyclic behaviour that will be explored in the next section.

\subsubsection*{Cyclic behaviour}

For very low values of $r$ and $\delta$, we observe that a cyclic behaviour is possible, where $ADD>ACD>NDD>ADD$ (here $X> Y$ or $Y < X$ means X invades Y).
One interesting observation is that all strategies involved defect if the commitment is not formed. This reinforces the idea that cooperating when a commitment is not formed is disadvantageous since it indicates the players' intention to cooperate is unlikely.
Each step of the cycle $ADD>ACD>NDD>ADD$ can be explained as follows.
For low enough $r$, $ADD$  invades $ACD$ while for high values of $r$, the opposite occurs. It is because the commitment is always formed and thus the former  defects in the interactions. 
Next, $ACD$ players  invade $NDD$ ones. Both defect if the commitment is not being formed. Despite that, $ACD$ has spatial reciprocity while $NDD$ doesn't. Thus, even if they defect when interacting with each other the $ACD$ players  have higher payoffs  as  they contribute in the groups where there are only $ACD$ players.
Now, $NDD$ players  invade $ADD$ ones, due to the fact that the latter are always being punished when a commitment is formed, while the former are not.

The cycle also occurs when replacing $NDD$ with $NCD$, where the same explanation applies. Since the cycle has a $ADD$ strategy, for a strong enough punishment, they become extinct, and only $ACD$  survive, breaking the cycle. 
However, if the punishment is sufficiently low, the invasion $NDD>ADD$ is  slow, which is detrimental for $ACD$ players' success and results in the dominance of defection. 
The cycles can happen for a variety of $r$ values but with a negligible  frequency of cooperation. The most relevant regions are for extremely low $r$ values if $\tau = 5$ and for small punishments for  $\tau = 3$ and $4$ (below the plateau). 

\subsection{ Reward ($\omega=1$, for varying $\delta$) }

\begin{figure}[H]
    \centering
    \includegraphics[width=0.3\columnwidth]{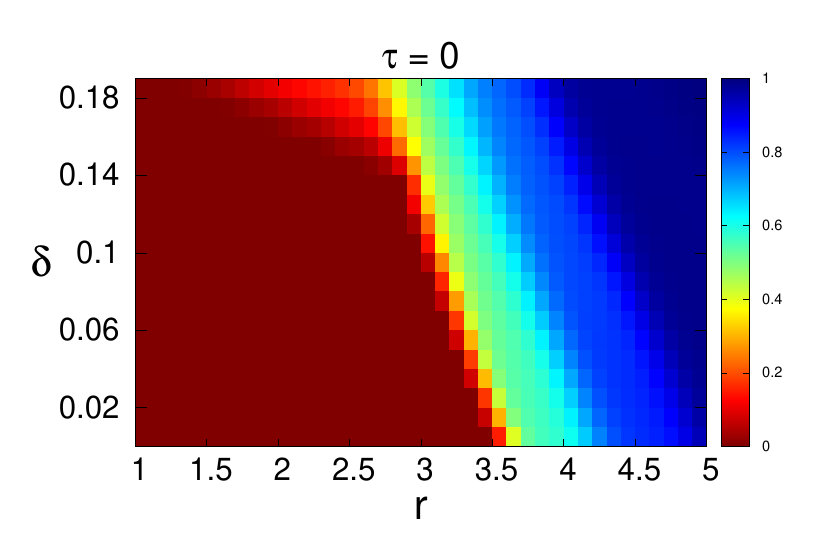}
    \includegraphics[width=0.3\columnwidth]{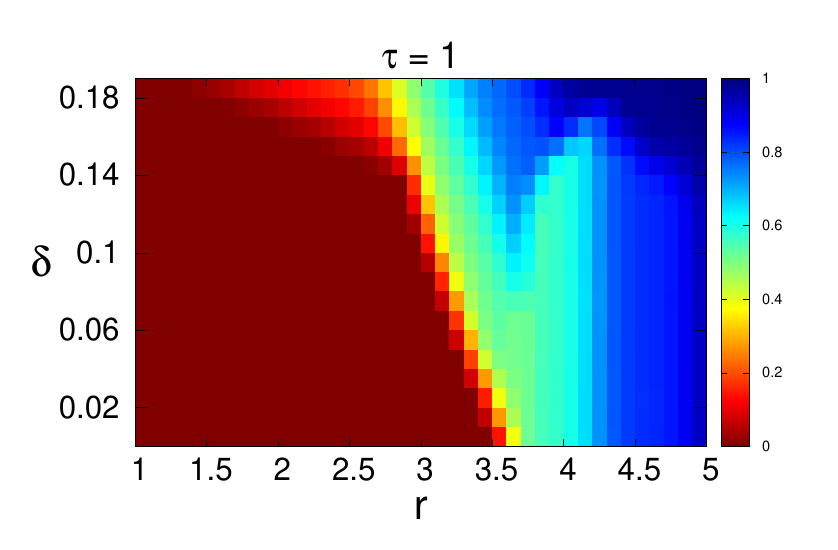}
    \includegraphics[width=0.3\columnwidth]{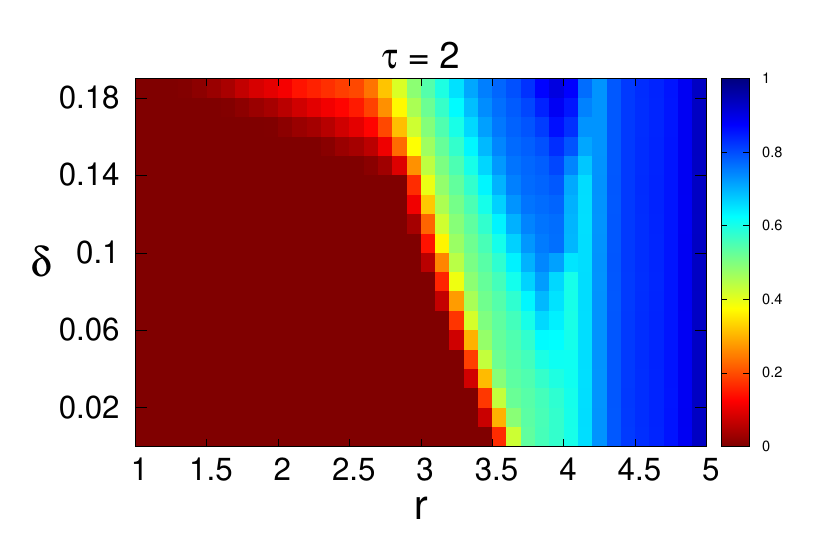}
    \includegraphics[width=0.3\columnwidth]{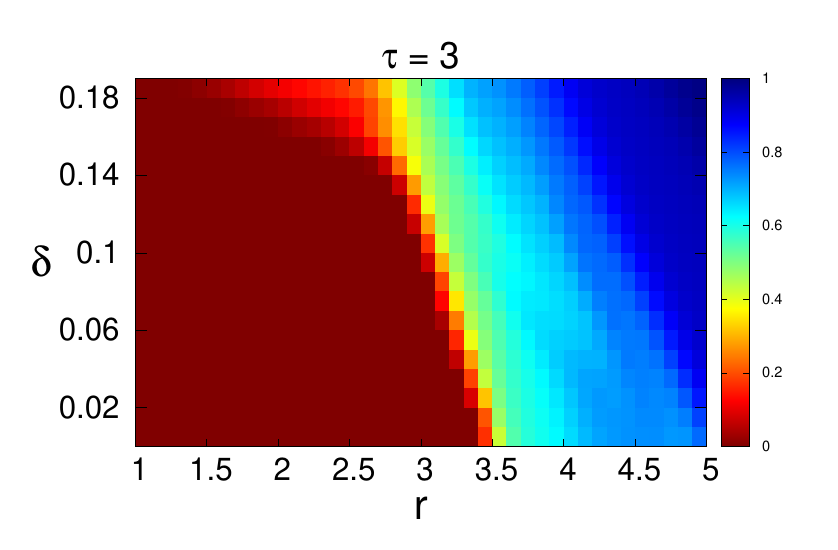}
    \includegraphics[width=0.3\columnwidth]{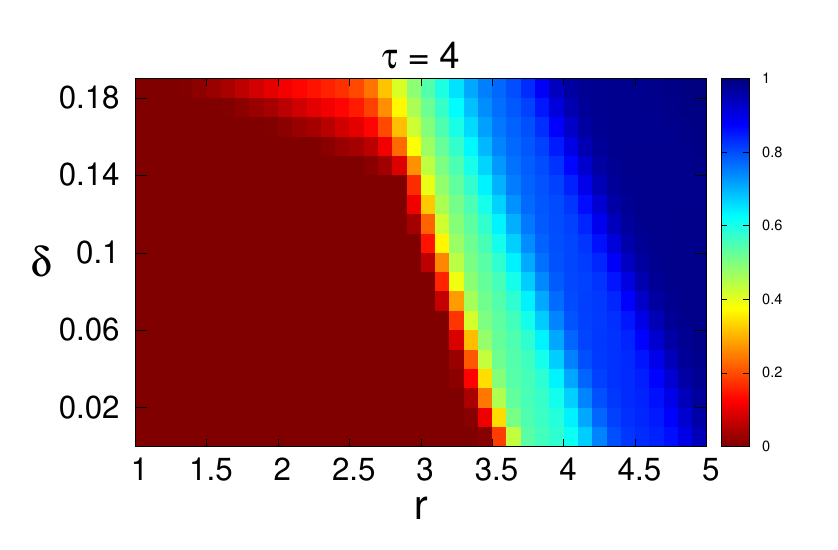}
    \includegraphics[width=0.3\columnwidth]{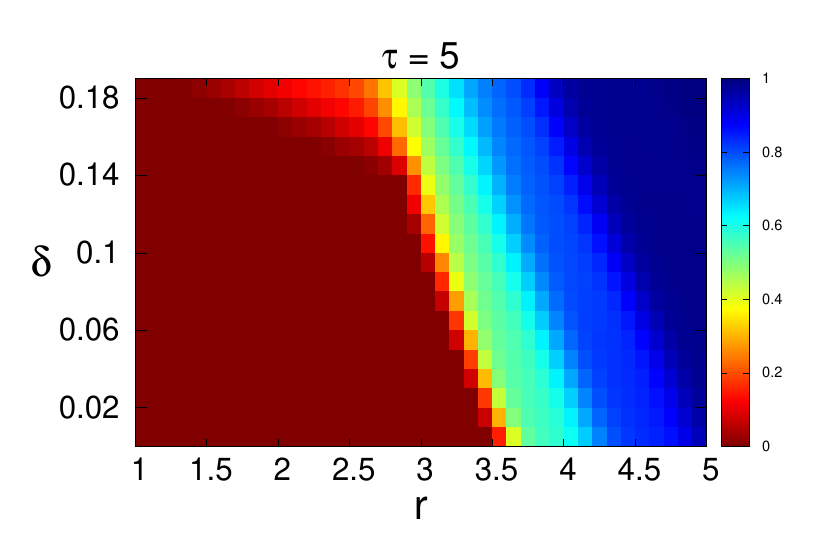}    
    \caption{Phase diagram $r \times \delta$ for the density of cooperative behaviour for all thresholds, in the presence of reward. We observe that for all thresholds, the critical values of $r$ for the survival of cooperation are similar. Despite that, some thresholds differ for high $r$ values.}
     \label{rewar}
\end{figure}

We now consider the situation where commitment-compliant players, i.e. those who commit to cooperate and actually cooperate when the commitment is formed, are rewarded. The rewarded players gain an increase in their payoff according to Table \ref{table:incentives}.
In Fig.~\ref{rewar}, we show the phase diagrams $r \times \delta$ of cooperative intensity for varying the threshold $\tau$. We observe that the effect of reward is almost invariant among the threshold values. 
Recall that in the absence of incentives (Section A), low thresholds selected for non-committers due to $NDC$ clustering and exploiting the other clustering strategies (namely, $NCC$, $ACC$, and $ACD$). Now, in the presence of reward, $ACD$ and $ACC$ are rewarded for committing and cooperating, thus even when being exploited by $NDC$ strategy, they can  prevail.

\begin{figure}[H]
    \centering
    \includegraphics[width=0.4\columnwidth]{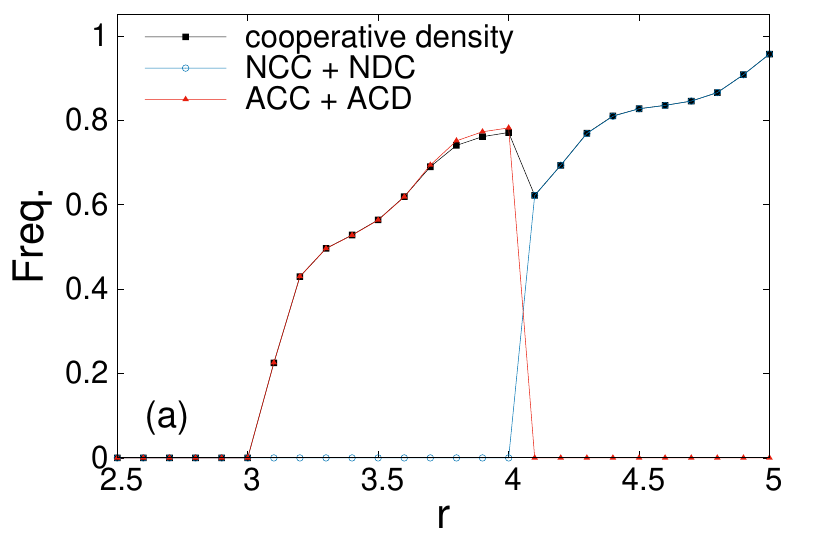} 
    \includegraphics[width=0.4\columnwidth]{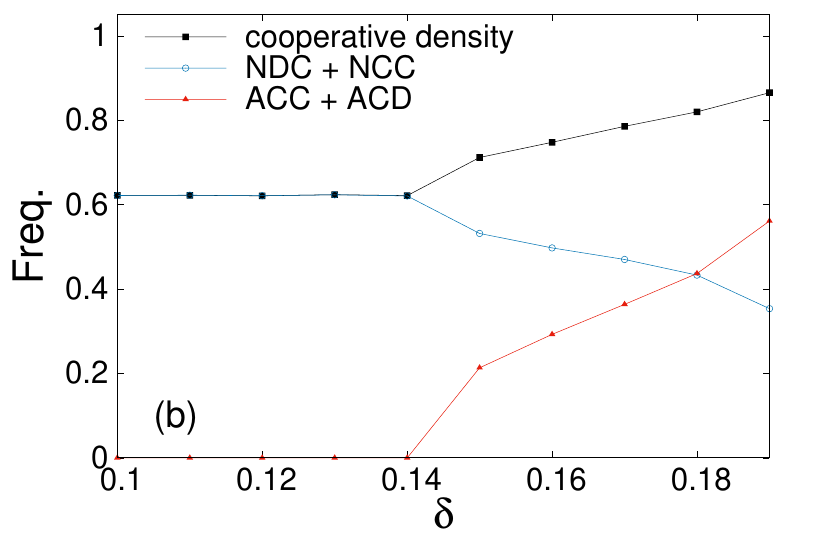}
    \caption{ Frequency of cooperation, committers ($ACC+ACD$) and non-committers ($NCC+NDC$) for $\tau=2$, as functions of:  $r$ (Left panel, a), for $\delta=0.1$, and  $\delta$ (Right panel, b), for $r=4.1$. In the former, similar to the behaviour discussed in the punishment case for $\tau=3$, we observe that for a high enough $r$ value non-committers can survive, thereby jeopardizing cooperation. For the latter (right panel), we observe that for a sufficiently high  reward value, committers can be favored where non-committers would have an advantage.    
    }
     \label{rewar2}
\end{figure}

We show, in Fig.~\ref{rewar2} (a), the density of cooperative committing ($ACC+ACD$) and non-committing ($NCC+NDC$) strategies for varying $r$ (fixing $\tau=2$ and $\delta=0.1$).
An interesting dynamics takes place now, where for $r<4$ only committers survive, including defectors; while for $r>4$, the population transitions into the dominance of non-committing strategies only.
This happens due to the fact that for low thresholds, $NDC$ can exploit and invade $ACD$. But if $r$ is low enough, non-committing cooperators cannot survive in the sea of defectors, since we would recover the classic PGG. Therefore, for low incentives there occurs a similar phase diagram to the punishment case for $\tau = 1$ and $2$, while  increasing the reward results in a new region of survival of cooperation due to committers.
This transition between committing and not committing ends up harming cooperation momentarily since cooperators were being rewarded but it ended for high $r$. 

Now, in Fig. \ref{rewar2} (b), we show the cooperative densities as function of $\delta$ (fixing $r=4.1$ and $\tau=2$). 
We observe that for low rewards we cannot have coexistence between committers and non-committers, as shown in (a).
But for sufficiently high reward values the committing cooperators are selected for and outperform the non-committing ones while they coexist. 
A sufficiently high reward benefits committers and therefore cooperation for low $\tau$. For higher thresholds, committers are selected for even without a reward, as previously explained.

Overall, given the analyses  above,  we arrive, first of all,  at the  conclusion that reward is preferred to punishment for promoting cooperation when the commitment threshold is low. 
When it is high,  punishment and reward have a similar effect on the cooperation outcomes. 
Nevertheless, there exists  clear difference between punishment and reward, which is most evident for $\tau=4$, where punishment ensures a higher advantage to cooperation.
Another notable finding from our analyses is that conditional strategies were essential for the maintenance of cooperation under punishment.
This is due to the fact that punishment does not necessarily mean a benefit to cooperators since non-committing defectors could still exist.
For the reward case, unconditional cooperators become more viable if the reward is high enough.
Finally, another noticeable finding is that $ADC$ and $NCD$ strategies are the least favorable in general. They can be deemed irrational or contradictory  since there is no clear benefit to commit and only cooperate if the commitment is not met. Despite that, $NCD$ could still survive  in the   punishment case due to possible patterns of cyclic dominance.

\section{Discussion}
In this paper, we have explored evolutionary dynamics in the spatial PGG  with the introduction of commitment.
In this setting, behavioural strategies can be conditional,  choosing cooperation or defection based on the  level of commitment from the group members.
Interestingly, with only this assumption, we found an increase in cooperation for a specific commitment threshold. 
This happens because cooperators that can change to defection in some situations outperform unconditional cooperators
In this case, we also found that having low commitment thresholds selected for only  non-committing strategies, while having high commitment thresholds selected for only  committing ones. 

We have also studied the effect of punishing commitment non-compliant players, showing that intermediate commitment thresholds required the least severe punishment  to sustain cooperative behaviour.
Nevertheless, to achieve highest levels of cooperation, it is necessary to strictly impose high commitment thresholds and strong punishment.

When considering the effect of rewarding commitment-compliant players, we found that all commitment thresholds led to similar cooperation outcomes. This is because, for any threshold, a high enough reward could sustain cooperators that committed. This was not the case with punishment since defectors could still avoid punishment by not committing.
The insight gleaned from this finding provides valuable implications for designing institutional mechanisms that foster pro-social behavior, particularly in situations where pre-communication is permitted to establish mutual  cooperative agreements \citep{chen1994effects, cherry2013enforcing, nesse2001evolution,NBOgbo,han2022voluntary}.

It is noteworthy that institutional incentives have been studied as an important pathway for promoting the emergence of cooperation in social dilemma situations, both in well-mixed and spatial settings   \citep{sasaki2012take,sigmund2010social,chen2015first,DuongHanPROCsA2021,CIMPEANU2021107545,gois2019reward,sun2021combination,boyd2010coordinated,hilbe2014democratic,flores2021symbiotic}. However, they have not  explored the repercussions of introducing a commitment before interactions in networks, a common occurrence in real-world personal and business settings \cite{barrett2003environment,nesse2001evolution,HanBook2013,frank88,cherry2013enforcing,balliet2010communication}.
Moreover, an issue with pro-social incentives aimed at promoting cooperation is the potential for antisocial reward and punishment dynamics. In this scenario, defectors may choose to punish cooperators or reward fellow defectors, thereby impeding the evolutionary progress of cooperation \cite{Herrmann2008,Rand2011,dos2015evolution}. 
This concern diminishes when a prior commitment is established, as it clarifies the expected behaviour of all parties involved in the interaction. As such, only those who commit to cooperation can face repercussions for defection or receive rewards for cooperation \citep{han2022Interface}. 

Furthermore, many previous evolutionary game  models have shown that different  heterogeneous aspects, including but not limited to network structures, update mechanisms and incentives,  play an important role for the emergence of cooperation \cite{perc2015double,attila_heterog, attila_heterog2, marco_heterog, CAO20101273,santos2008social,flores2023heterogeneous,perc2017statistical}. %These heterogeneous factors can 
As a future work, it would be interesting to incorporate heterogeneity into our model. This would specifically involve examining the influence of individualised commitment thresholds, for each player, rather than treating them as global parameters; as well as  that of different heterogeneous  population structures.

\section*{CRediT authorship contribution statement}

L. S. Flores contributed with Software and all authors contributed equally in Conceptualization,
Formal analysis and Writing.

 \section*{Acknowledgments}
L.S.Flores thanks the Brazilian funding agency CAPES (Coordenação de Aperfeiçoamento de Pessoal de Nível Superior) for the Ph.D. scholarship. The simulations were performed on the IF-UFRGS computing cluster infrastructure.

\bibliographystyle{abbrv}
\bibliography{refs} % replace by the name of your .bib file

\end{document}